\documentclass[prb,10pt,superscriptaddress,showpacs,amsmath,twocolumn,floatfix]{revtex4}
\usepackage{amsmath}
\usepackage{times}
\usepackage{graphicx}
\usepackage{mathptmx}

\newcommand{\parti}[2]{\frac{\partial #1}{\partial #2}}

\newcommand{\opex}{Opt.\ Express }

\begin{document}
%\twocolumn[
\title{Magnifying perfect lens and superlens design by coordinate
transformation}

\author{Mankei Tsang}
\email{mankei@optics.caltech.edu}
\affiliation{Department of Electrical Engineering, 
California Institute of Technology, Pasadena, California 91125, USA}
\author{Demetri Psaltis}
\affiliation{Department of Electrical Engineering, 
California Institute of Technology, Pasadena, California 91125, USA}
\affiliation{Institute of Imaging and Applied Optics,
Ecole Polytechnique F\'ed\'erale de Lausanne,
CH-1015 Lausanne, Switzerland}

\date{\today}

\begin{abstract}
  The coordinate transformation technique is applied to the design of
  perfect lenses and superlenses. In particular, anisotropic
  metamaterials that magnify two-dimensional planar images beyond the
  diffraction limit are designed by the use of oblate spheroidal
  coordinates. The oblate spheroidal perfect lens or superlens can
  naturally be used in reverse for lithography of planar subwavelength
  patterns.
\end{abstract}
%\ocis{(160.3918)   Metamaterials;
%(180.4243) Near-field microscopy;
%(110.3960) Microlithography;
%(240.6680) Surface plasmons;
%(250.5403) Plasmonics
%(100.6640) Superresolution;
%}
%]
\maketitle

\section{Introduction}
Leonhardt \cite{leonhardt} and Pendry \textit{et al.}\ \cite{pendry}
recently suggested an interesting technique of controlling
the propagation of electromagnetic fields by the
use of metamaterials. In this paper we shall apply this technique to
the design of perfect lenses \cite{veselago,pendry_prl,pendry_oe},
which are able to perfectly reproduce an image on another surface, and
superlenses
\cite{pendry_prl,fang,melville,salandrino,jacob,liu,smolyaninov},
which apply only to transverse-magnetic (TM) waves. In particular, we
show that the technique can be used to design transformation media
that magnify images beyond the diffraction limit. Perfect cylindrical
lenses have been proposed by Pendry \cite{pendry_oe}, while
cylindrical magnifying superlenses were recently proposed by
Salandrino and Engheta \cite{salandrino} and Jacob \textit{et al.}\
\cite{jacob} and experimentally demonstrated by Liu \textit{et al.}\
\cite{liu} and Smolyaninov \textit{et al.}\ \cite{smolyaninov}. We
show that the principle behind such cylindrical devices can be
generalized to arbitrary three-dimensional orthogonal coordinate
systems.  Using the oblate spheroidal coordinates, we further show how
perfect lenses and superlenses that magnify planar images with
subwavelength features can be designed.  The flat object plane is more
convenient for imaging and lithography applications.

Our approach yields fundamentally different results from the brief
discussion on magnifying perfect lenses in
Ref.~\onlinecite{leonhardt_njp}. We discuss this discrepancy in
Section \ref{planar} and argue that the perfect lens design proposed
in Ref.~\onlinecite{leonhardt_njp} does not provide magnification, but
rather changes the depth of field or depth of focus only. Our
magnifying superlens design, outlined in Sec.~\ref{superlens}, is also
more general and different than that in
Ref.~\onlinecite{salandrino}, in order to avoid the problem of
impedance mismatch between the metamaterial with zero transverse
permittivity and free space.

\section{Maxwell Equations under Coordinate Transformation}
\begin{figure}[htbp]
\centerline{\includegraphics[width=0.48\textwidth]{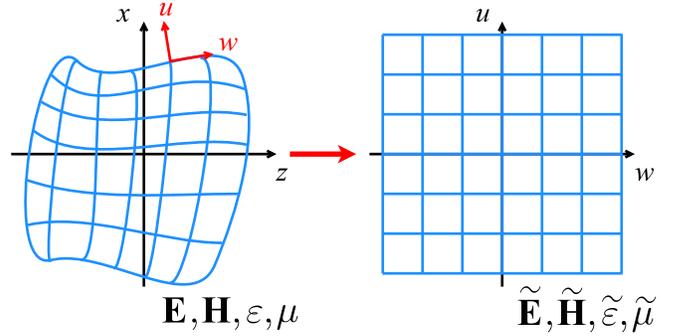}}
\caption{(Color online) Electromagnetic fields $\mathbf{E}$ and
  $\mathbf{H}$ in a physical medium with material constants $\epsilon$
  and $\mu$ can be transformed to normalized fields
  $\tilde{\mathbf{E}}$ and $\tilde{\mathbf{H}}$ that also obey the
  Maxwell equations, but regard $(u(x,y,z),v(x,y,z),w(x,y,z))$ as
  Cartesian coordinates in an effective medium with material constants
  $\tilde{\epsilon}$ and $\tilde{\mu}$.}
\label{conformal}
\end{figure}

For completeness and to establish our notations, we shall first
briefly review the invariant property of the Maxwell equations under
an orthogonal coordinate transformation \cite{moon}, closely following
Pendry \textit{et al.}\ \cite{pendry,ward}.  The Maxwell equations in
terms of harmonic fields in Cartesian coordinates $(x,y,z)$ are
\begin{align}
\nabla\cdot \left(\epsilon\mathbf{E}\right) &= 0,
&
\nabla\cdot \left(\mu\mathbf{H}\right) &= 0,
\nonumber\\
\nabla\times\mathbf{E} &=i\omega\mu_0\mu\mathbf{H},
&
\nabla\times\mathbf{H} &= -i\omega \epsilon_0\epsilon\mathbf{E},
\end{align}
where both $\epsilon$ and $\mu$ are second-rank tensors.
With a new set of orthogonal coordinates $(u,v,w)$,
\begin{align}
u &= u(x,y,z), & v &= v(x,y,z), & w &= w(x,y,z),
\nonumber\\
x &= x(u,v,w), & y &= y(u,v,w), & z &= z(u,v,w),
\end{align}
the fields and the material constants can be rewritten in terms of the
new coordinates as
\begin{align}
\mathbf{E}(u,v,w) &\equiv \mathbf{E}\left[x(u,v,w),y(u,v,w),z(u,v,w)\right]
\nonumber\\
&=\sum_{q=u,v,w} E_q(u,v,w) \hat{\mathbf{q}},
\nonumber\\
\epsilon(u,v,w) &\equiv
\epsilon\left[x(u,v,w),y(u,v,w),z(u,v,w)\right]
\nonumber\\
&=\sum_{p,q}\epsilon_{pq}(u,v,w) \hat{\mathbf{p}}\hat{\mathbf{q}},
\end{align}
and likewise for $\mathbf{H}$ and $\mu$.  We assume that $\epsilon$
and $\mu$ are diagonal in the new coordinates, such that
\begin{align}
\epsilon_{pq} &= \epsilon_{q}\delta_{pq},
& \mu_{pq} &= \mu_{q}\delta_{pq}.
\end{align}
If we define the following normalized fields and material constants,
\begin{align}
\left(\tilde{E}_u,\tilde{E}_v,\tilde{E}_w\right) &\equiv
\left(h_u E_u, h_vE_v, h_wE_w\right),
\nonumber\\
\left(\tilde{H}_u,\tilde{H}_v,\tilde{H}_w\right) &\equiv
\left(h_u H_u, h_vH_v, h_wH_w\right),
\nonumber\\
\left(\tilde{\epsilon}_u,\tilde{\epsilon}_v,\tilde{\epsilon}_w\right)
&\equiv h_uh_vh_w
\left(\frac{\epsilon_u}{h_u^2},\frac{\epsilon_v}{h_v^2},\frac{\epsilon_w}{h_w^2}\right),
\nonumber\\
\left(\tilde{\mu}_u,\tilde{\mu}_v,\tilde{\mu}_w\right)
&\equiv h_uh_vh_w
\left(\frac{\mu_u}{h_u^2},\frac{\mu_v}{h_v^2},\frac{\mu_w}{h_w^2}\right),
\label{norm_quantities}
\end{align}
where $h_q$ is the scale factor of the new coordinates \cite{arfken},
also called the Lam\'e coefficients \cite{lame},
\begin{align}
h_q(u,v,w) = \sqrt{\left(\parti{x}{q}\right)^2+
\left(\parti{y}{q}\right)^2+\left(\parti{z}{q}\right)^2},
\label{scale}
\end{align}
the normalized quantities $\tilde{\mathbf{E}}$, $\tilde{\mathbf{H}}$,
$\tilde{\epsilon}$, and $\tilde{\mu}$ satisfy the same Maxwell
equations, but now they see $(u,v,w)$ as Cartesian coordinates,
\begin{align}
\tilde\nabla\cdot(\tilde\epsilon\tilde{\mathbf{E}}) &= 0,
&
\tilde\nabla\cdot \left(\tilde\mu\tilde{\mathbf{H}}\right) &= 0,
\nonumber\\
\tilde{\nabla}\times\tilde{\mathbf{E}}&=
i\omega\mu_0\tilde\mu\tilde{\mathbf{H}},
&
\tilde{\nabla}\times\tilde{\mathbf{H}} &= -i\omega\epsilon_0\tilde\epsilon
\tilde{\mathbf{E}},
\label{norm_maxwell}
\end{align}
where
\begin{align}
\tilde\nabla &\equiv \hat{\mathbf{u}} \parti{}{u} + \hat{\mathbf{v}}\parti{}{v}
+\hat{\mathbf{w}}\parti{}{w}.
\end{align}
In other words, the electromagnetic fields $\mathbf{E}$ and
$\mathbf{H}$ in a physical medium with material constants $\epsilon$
and $\mu$ can be transformed to normalized fields $\tilde{\mathbf{E}}$
and $\tilde{\mathbf{H}}$ that also obey the Maxwell equations, but
regard $(u(x,y,z),v(x,y,z),w(x,y,z))$ as Cartesian coordinates in an
effective medium with material constants $\tilde\epsilon$ and
$\tilde\mu$.  See Fig.~\ref{conformal} for an illustration of the
coordinate transformation.

\section{\label{perfect_design}Perfect Lens Design}
\subsection{General Procedure}
In general, a perfect lens should transmit the electromagnetic fields
from one surface to another surface with perfect fidelity and without
any reflection \cite{pendry_prl}.  Let us define a \emph{physical}
space with a coordinate system $(u',v',w')$ that represent the two
surfaces by the equations $w'(x',y',z') = a$ and $w'(x',y',z') = b$,
respectively. If we fill the volume between these two surfaces with
metamaterial, an appropriate design of the metamaterial can map the
actual fields on these two surfaces onto any other pair of surfaces in
a \emph{virtual} space \cite{leonhardt_njp}, with another coordinate
system $(u,v,w)$, so that the fields propagate in a physical medium
with material constants $(\epsilon_{u'},\epsilon_{v'}, \epsilon_{w'})$
and $(\mu_{u'},\mu_{v'}, \mu_{w'})$ across two surfaces in the
physical space as if they propagate across the two mapped surfaces in
a virtual medium with $(\epsilon_{u},\epsilon_{v}, \epsilon_{w})$ and
$(\mu_{u},\mu_{v}, \mu_{w})$.

\begin{figure}[htbp]
\centerline{\includegraphics[width=0.48\textwidth]{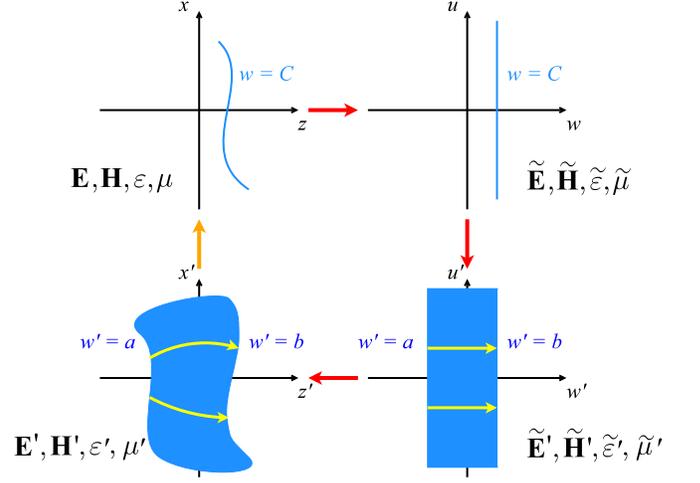}}
\caption{(Color online) The procedure of perfect lens design by the
  coordinate transformation technique. First a curved surface $w = C$
  in the virtual space (top left) is transformed into a plane (top
  right). The plane is then mapped onto a slab (bottom right), which
  is subsequently transformed back to the desired geometry in the
  physical space (bottom left). The electromagnetic fields propagate
  through the transformation medium with material constants
  $\epsilon'$ and $\mu'$ in the physical space as if they propagate
  within an infinitesimal slab in the virtual space. The fields from a
  point source on the $w'=a$ surface propagate like a ray, depicted by
  yellow arrows, along a $w'$ coordinate line inside the
  transformation medium.}
\label{perfect_lens}
\end{figure}

To make the fields propagate from $w' = a$ to $w' = b$ without any
distortion, one can map the two surfaces in the physical space onto
the same surface in the virtual space.  A straightforward way is to
map \emph{all} constant-$w'$ surfaces within $a \le w' \le b$ in the
physical space onto a single constant-$w$ surface in the virtual
space. Mathematically, such a mapping can be achieved if
\begin{align}
u &= u',& v &= v', & w &= C, &  a\le w' \le b,
\label{transform}
\end{align}
where $C$ is an arbitrary constant. The corresponding scale factors
are
\begin{align}
\tilde{h}_{q'} \equiv
\sqrt{\left(\parti{u}{q'}\right)^2+\left(\parti{v}{q'}\right)^2
+\left(\parti{w}{q'}\right)^2}.
\label{tildescale}
\end{align}
The surface mapping function $w = C$ can clearly be generalized to
accomodate other requirements. For example, a negative-index material
can be used to produce a negative mapping function $w = -w'$, so that
part of the perfect lens can be free space and the working distance
can be increased \cite{leonhardt_njp}. Nonetheless, in the following
we shall use the constant mapping $w = C$ for simplicity. The
transformation in the other regions ($w' < a, w' > b$) can be
exploited to simplify the lens design.

To design the metamaterial properties, we should first specify the
target virtual material properties
$(\epsilon_u,\epsilon_v,\epsilon_w)$ and $(\mu_u,\mu_v,\mu_w)$. For
example, if we want the fields to propagate in a virtual free space,
we should set $(\epsilon_u,\epsilon_v,\epsilon_w) =
(\mu_u,\mu_v,\mu_w) = (1,1,1)$. Next, we transform the fields and
material constants in the virtual space to normalized ones,
\begin{align}
\left(\tilde{E}_u,\tilde{E}_v,\tilde{E}_w\right) &=
\left(h_u E_u, h_vE_v, h_wE_w\right),
\nonumber\\
\left(\tilde{H}_u,\tilde{H}_v,\tilde{H}_w\right) &=
\left(h_u H_u, h_vH_v, h_wH_w\right),
\nonumber\\
\left(\tilde{\epsilon}_u,\tilde{\epsilon}_v,\tilde{\epsilon}_w\right)
&= h_uh_vh_w
\left(\frac{\epsilon_u}{h_u^2},\frac{\epsilon_v}{h_v^2},\frac{\epsilon_w}{h_w^2}\right),
\nonumber\\
\left(\tilde{\mu}_u,\tilde{\mu}_v,\tilde{\mu}_w\right)
&= h_uh_vh_w
\left(\frac{\mu_u}{h_u^2},\frac{\mu_v}{h_v^2},\frac{\mu_w}{h_w^2}\right),
\end{align}
so that $(u,v,w)$ become the new Cartesian coordinates in a normalized
virtual space. With the coordinate transformation from $(u,v,w)$ to
$(u',v',w')$ specified by Eqs.~(\ref{transform}) that maps
the $w = C$ surface to all constant-$w'$ surfaces in the normalized
physical space, the normalized quantities become
\begin{align}
\left(\tilde{E}_{u'},\tilde{E}_{v'},\tilde{E}_{w'}\right) &=
\left(\tilde{h}_{u'}\tilde{E}_u, \tilde{h}_{v'}\tilde{E}_v,\tilde{h}_{w'}\tilde{E}_w\right),
\nonumber\\
\left(\tilde{H}_{u'},\tilde{H}_{v'},\tilde{H}_{w'}\right) &=
\left(\tilde{h}_{u'}\tilde{H}_u, \tilde{h}_{v'}\tilde{H}_v,\tilde{h}_{w'}\tilde{H}_w\right),
\nonumber\\
\left(\tilde{\epsilon}_{u'},\tilde{\epsilon}_{v'},\tilde{\epsilon}_{w'}\right)
&= \tilde{h}_{u'}\tilde{h}_{v'}\tilde{h}_{w'}
\left(\frac{\tilde\epsilon_u}{\tilde{h}_{u'}^2},
\frac{\tilde\epsilon_v}{\tilde{h}_{v'}^2},\frac{\tilde\epsilon_w}{\tilde{h}_{w'}^2}\right),
\nonumber\\
\left(\tilde{\mu}_{u'},\tilde{\mu}_{v'},\tilde{\mu}_{w'}\right)
&= \tilde{h}_{u'}\tilde{h}_{v'}\tilde{h}_{w'}
\left(\frac{\tilde\mu_u}{\tilde{h}_{u'}^2},
\frac{\tilde\mu_v}{\tilde{h}_{v'}^2},\frac{\tilde\mu_w}{\tilde{h}_{w'}^2}\right),
\end{align}
Since these new quantities see $(u',v',w')$ as Cartesian coordinates,
we should perform the inverse coordinate transform on the normalized
quantities, in order to obtain the physical ones that regard
$(u',v',w')$ as the desired non-Cartesian coordinate system,
\begin{align}
\left(E_{u'},E_{v'},E_{w'}\right) &=
\left(\frac{\tilde{E}_{u'}}{h_{u'}}, \frac{\tilde{E}_{v'}}{h_{v'}},
\frac{\tilde{E}_{w'}}{h_{w'}}\right),
\nonumber\\
\left(H_{u'},H_{v'},H_{w'}\right) &=
\left(\frac{\tilde{H}_{u'}}{h_{u'}}, \frac{\tilde{H}_{v'}}{h_{v'}},
\frac{\tilde{H}_{w'}}{h_{w'}}\right),
\nonumber\\
\left(\epsilon_{u'},\epsilon_{v'},\epsilon_{w'}\right)
&=\frac{1}{h_{u'}h_{v'}h_{w'}}
\left(h_{u'}^2\tilde{\epsilon}_u,h_{v'}^2\tilde{\epsilon}_v,
h_{w'}^2\tilde{\epsilon}_w\right),
\nonumber\\
\left(\mu_{u'},\mu_{v'},\mu_{w'}\right)
&=\frac{1}{h_{u'}h_{v'}h_{w'}}
\left(h_{u'}^2\tilde{\mu}_u,h_{v'}^2\tilde{\mu}_v,
h_{w'}^2\tilde{\mu}_w\right),
\end{align}
where $h_{q'}$ is the same scale factor defined in Eq.~(\ref{scale}),
but with $q' = u',v',w'$. As all the constant-$w'$ surfaces in the
physical space are mapped onto the same surface in the virtual space
and thus have identical normalized fields, the fields from a point
source on the $w' = a$ surface propagate like a ray inside the
transformation medium. The rays follow the $w'$ coordinate lines,
defined as lines along which $u'$ and $v'$ are constant, much like the
rays in an anisotropic metamaterial crystal described by Salandrino
and Engheta \cite{salandrino}. As the coordinate transformation
technique is applied to the full Maxwell equations, it also guarantees
that waves of arbitrary polarizations can be transmitted perfectly by
a perfect lens. Figure \ref{perfect_lens} depicts the procedure of
perfect lens design by the coordinate transformation technique
outlined above.

\subsection{\label{planar}Planar Perfect Lens}
The simplest example is planar imaging with no magnification. One can
use Cartesian coordinates for $(u,v,w)$ and $(u',v',w')$ and
the following transformation:
\begin{align}
x &= x', & y &= y', & z &= 
\bigg\{\begin{array}{l}
z'+b, \\
\delta z'+b,\\
z'+\delta b
\end{array}
&
\begin{array}{l}
z' < 0,\\
0 \le z' \le b,\\
z' > b,
\end{array}
\nonumber\\
\tilde{h}_{x'} &= 1, &
\tilde{h}_{y'} &=1, &
\tilde{h}_{z'} &=
\bigg\{\begin{array}{l}
1,\\
\delta,\\
1,
\end{array}
&
\begin{array}{l}
z' < 0,\\
0 \le z' \le b,\\
z' > b,
\end{array}
\end{align}
and let $\delta\to 0$ at the end of the calculation.  Assuming that
the virtual space is free space, such that
$(\epsilon_x,\epsilon_y,\epsilon_z) = (\mu_x,\mu_y,\mu_z) = (1,1,1)$,
and using the procedure outlined above, we obtain the following
desired physical material constants:
\begin{align}
\left(\epsilon_{x'},\epsilon_{y'},\epsilon_{z'}\right)
&=\left(\mu_{x'},\mu_{y'},\mu_{z'}\right)
=\Big\{\begin{array}{l}
\left(\delta,\delta,1/\delta\right),\\
\left(1,1,1\right), 
\end{array}
&
\begin{array}{l}
 0 \le z' \le b,\\
\textrm{otherwise.}
\end{array}
\label{perfect_slab}
\end{align}
The slab is a perfectly matched layer \cite{berenger}, as one would
expect for a reflectionless structure. In the limit of $\delta\to 0$,
the fields propagate in the metamaterial slab as if they propagate in
an infinitesimal slab at $z = b$ in the virtual free space, so that
the fields on one side ($z' = 0^-$) are perfectly transmitted to the
other ($z' = b^+$) without any reflection.

In Ref.~\onlinecite{leonhardt_njp}, the authors assert that a magnifying
perfect lens can be achieved if $\delta$ is negative and different
from 1. Their approach yields the following coordinate transformation:
\begin{align}
x &= x', & y &= y', & z &= -|\delta|z'+C, & 0 \le z' \le b.
\end{align}
This coordinate transformation clearly does not provide any
magnification, as the transverse coordinates $x$ and $y$ are
unchanged, but rather it changes the depth of field or depth of focus
only. Instead of producing a magnified perfect image, a misplaced
depth of field or depth of focus can only blur the image on the
desired image plane, or reproduce a non-magnified perfect image on a
different plane.

\subsection{\label{implement}Metamaterial Implementation}
\begin{figure}[htbp]
\centerline{\includegraphics[width=0.48\textwidth]{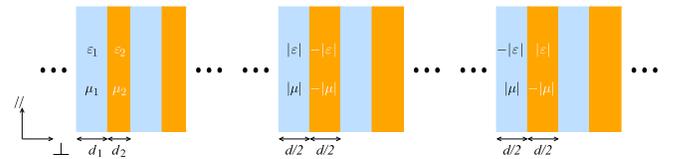}}
\caption{(Color online) Effective anisotropic metamaterial formed by
  thin films (left), and two possible realizations of the anistropic
  perfect lens (center and right).  The perfect lens can be formed by
  pairing positive-refractive-index films with
  negative-refractive-index films of the same thickness (center), as
  suggested by Veselago \cite{veselago} and Pendry \cite{pendry_prl},
  or pairing negative-$\epsilon$ films with negative-$\mu$ films
  (right), as suggested by Al\`u and Engheta \cite{alu}.  }
\label{stack}
\end{figure}

The highly anisotropic metamaterial specified by
Eqs.~(\ref{perfect_slab}) in the limit of $\delta\to 0$ can be
approximately implemented by a stack of thin slabs with alternate
signs of permittivity and permeability
\cite{rytov,alu,ramakrishna,schurig,tretyakov} (Fig.~\ref{stack}). It
can be shown, by generalizing the argument in
Ref.~\onlinecite{tretyakov}, that the effective material constants of
the stack shown in the left figure of Fig.~\ref{stack} in the limit of
$d_1 \ll \lambda/|n_1|$ and $d_2 \ll \lambda/|n_2|$ are
\begin{align}
\epsilon_{\parallel} &\approx\frac{\epsilon_1 d_1 + \epsilon_2 d_2}{d_1+d_2},
&
\epsilon_{\perp} &\approx\frac{d_1+d_2}{d_1/\epsilon_1+d_2/\epsilon_2},
\nonumber\\
\mu_{\parallel} &\approx \frac{\mu_1 d_1 + \mu_2 d_2}{d_1+d_2},
&
\mu_{\perp} &\approx \frac{d_1+d_2}{d_1/\mu_1+d_2/\mu_2},
\end{align}
With $\epsilon_1 = -\epsilon_2$ and $\mu_1 = -\mu_2$, the desired
anisotropic metamaterial properties are achieved. Two different
possible realizations are sketched in the center and right figures of
Fig.~\ref{stack}.

We note that the effective medium theory is not exact if the
thicknesses of the layers are not significantly smaller than the
wavelength \cite{tretyakov,vinogradov}.  Deviation from Rytov's
effective medium theory of stratified media is an interesting subject
that deserves further investigation, but it is beyond the scope of
this paper to study this issue, and in the rest of the paper we shall
follow the effective medium theory as the first order approximation,
which has otherwise withstood theoretical, numerical, and experimental
tests \cite{salandrino,jacob,liu,smolyaninov,yariv}. Advance in
metamaterial technology may also enable new alternative
implementations of the desired material parameters.

\section{Magnifying Perfect Lenses}
\subsection{\label{spherical_lens}Spherical Perfect Lens}
\begin{figure}[htbp]
\centerline{\includegraphics[width=0.48\textwidth]{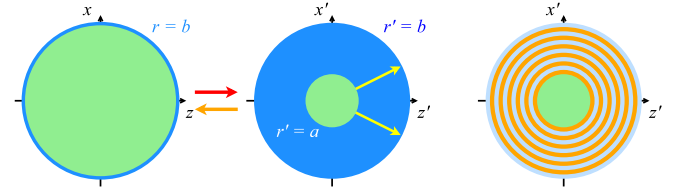}}
\caption{(Color online) The spherical magnifying perfect lens
  (center), the corresponding virtual space (left), and the
  metamaterial implementation of the lens (right). The corresponding
  regions in the virtual space and the physical space are marked by
  the same colors in the left and center figures. For clarity, only
  the $y= 0$ cross section is shown.  The electromagnetic fields on
  the inner spherical surface are perfectly mapped onto the outer
  surface by the lens, enabling far-field detection of subwavelength
  information. In practice, one can use just half of the spherical
  lens, so that the object can be placed against the inner spherical
  surface more conveniently.}
\label{spherical}
\end{figure}

To achieve magnification, one surface in the physical space, say $w' =
a$, can be defined to accomodate the object geometry, while the other
surface, $w' = b$, can be mapped to a larger area, thus converting the
fields to far-field radiation for easier detection. The magnifying
perfect lens can naturally be used in reverse for lithography.  One
coordinate system that can achieve magnification is the spherical
coordinate system, a natural three-dimensional generalization of the
cylindrical geometry studied in
Refs.~\onlinecite{pendry_oe,salandrino,jacob,liu,smolyaninov}:
\begin{align}
x &= r\sin\theta\cos\phi, &
y &= r\sin\theta\sin\phi, &
z &= r\cos\theta.
\nonumber\\
h_{\theta} &= r, &
h_{\phi} &= r\sin\theta, &
h_{r} &= 1.
\end{align}
We shall use the following coordinate transformation:
\begin{align}
\theta &= \theta', &
\phi &= \phi', &
r &= \bigg\{
\begin{array}{l}
br'/a,\\
b,  \\
r',
\end{array}
&
\begin{array}{l}
r' < a,\\
a \le r' \le b, \\
r' > b,
\end{array}
\nonumber\\
\tilde{h}_{\theta'} &= 1, &
\tilde{h}_{\phi'} &= 1, &
\tilde{h}_{r'} &= \bigg\{\begin{array}{l}
b/a,\\
0, \\
1,
\end{array}
&
\begin{array}{l}
r' < a,\\
a \le r' \le b, \\
r' > b,
\end{array}
\end{align}
so that all spherical surfaces with $a \le r' \le b$ are mapped onto a
single spherical surface $r = b$ in the virtual space. The coordinate
transformation procedure yields
\begin{align}
\left(\epsilon_{\theta'},\epsilon_{\phi'},\epsilon_{r'}\right)
&=\frac{b}{a}\left(\epsilon_{\theta},\epsilon_{\phi},\epsilon_{r}\right),
\nonumber\\
\left(\mu_{\theta'},\mu_{\phi'},\mu_{r'}\right)
&=\frac{b}{a}\left(\mu_{\theta},\mu_{\phi},\mu_{r}\right)
\label{spherical_transform1}
\end{align}
for $r'<a$,
\begin{align}
\left(\epsilon_{\theta'},\epsilon_{\phi'},\epsilon_{r'}\right)
&=\left(\mu_{\theta'},\mu_{\phi'},\mu_{r'}\right)
=\left(0,0,\infty\right)
\label{spherical_transform2}
\end{align}
for $a \le r' \le b$, and
\begin{align}
\left(\epsilon_{\theta'},\epsilon_{\phi'},\epsilon_{r'}\right)
&=\left(\mu_{\theta'},\mu_{\phi'},\mu_{r'}\right)
=\left(1,1,1\right)
\label{spherical_transform3}
\end{align}
for $r' > b$.
If we let the virtual space be free space, the desired
physical material constants become
\begin{align}
\left(\epsilon_{\theta'},\epsilon_{\phi'},\epsilon_{r'}\right)
&=\left(\mu_{\theta'},\mu_{\phi'},\mu_{r'}\right)
= \bigg\{\begin{array}{l}
\left(\frac{b}{a},\frac{b}{a},\frac{b}{a}\right),\\
(0,0,\infty),\\
(1,1,1),
\end{array}
&
\begin{array}{l}
r' < a,\\
a \le r' \le b,\\
r' > b.
\end{array}
\end{align}
The transformation medium consists of an isotropic high-permittivity and
high-permeability material for $r'<a$, a highly anistropic shell for
$a\le r' \le b$ that can again be implemented by layers of thin
spherical shells with alternate signs of permittivity and
permeability, and free space for $r'>b$. Figure \ref{spherical}
depicts the geometry of the spherical perfect lens, the corresponding
virtual space, and the metamaterial implementation.

The spherical object surface is assumed to be situated at
$r' = a$, and any electromagnetic fields on the object surface are
perfectly transmitted to the outer spherical surface without any
reflection. The fields at $r'=a^-$ are related to the fields at
$r' = b^+$ by
\begin{align}
\mathbf{E}(\theta',\phi',b^+)
&= \frac{a}{b} \mathbf{E}(\theta',\phi',a^-),
\nonumber\\
\mathbf{H}(\theta',\phi',b^+)
&= \frac{a}{b} \mathbf{H}(\theta',\phi',a^-).
\end{align}
For large $b$, the fields become primarily far-field radiation at the
outer spherical surface that can be detected by conventional far-field
optics.

If we make the inner sphere $r'<a$ empty for practical reasons, so
that $(\epsilon_{\theta'},\epsilon_{\phi'},\epsilon_{r'}) =
(\mu_{\theta'},\mu_{\phi'},\mu_{r'}) = (1,1,1)$ for $r' < a$, the
fields no longer see the whole virtual space as free space, but as a
low-refractive-index sphere with radius $r = b$,
\begin{align}
\left(\epsilon_{\theta},\epsilon_{\phi},\epsilon_{r}\right)
&=\left(\mu_{\theta},\mu_{\phi},\mu_{r}\right)
= \bigg\{\begin{array}{l}
\left(\frac{a}{b},\frac{a}{b},\frac{a}{b}\right),\\
(1,1,1),
\end{array}
&
\begin{array}{l}
r < b,\\
r \ge b,
\end{array}
\end{align}
which can be derived from Eqs.~(\ref{spherical_transform1}).  In this
case, although the fields on each spherical surface within the
metamaterial lens for $a \le r' \le b$ still have the same azimuthal
profiles and are perfectly matched to the outer free space, there is
reflection and partial transmission across the inner interface of the
metamaterial shell, just as there is reflection and partial
transmission across the $r = a$ interface in the virtual space. In
other words, there is impedance mismatch between an empty inner volume
and the spherical lens, but the image transmission is still
perfect. The effects of other deviations from the perfect lens design
can also be understood by making more general coordinate
transformations and studying the electromagnetic field propagation in
the virtual space.

We note that Ref.~\onlinecite{liu} mentions the possibility of a
spherical superlens, while Narimanov's group at Purdue University also
allegedly has unpublished work regarding a spherical superlens, but to
our knowledge this is the first time that the specification of a
spherical perfect lens and that of a spherical superlens, to be
discussed in Sec.~\ref{superlens}, are reported.

\subsection{\label{oblate_lens}Oblate Spheroidal Perfect Lens}
\begin{figure}[htbp]
\centerline{\includegraphics[width=0.4\textwidth]{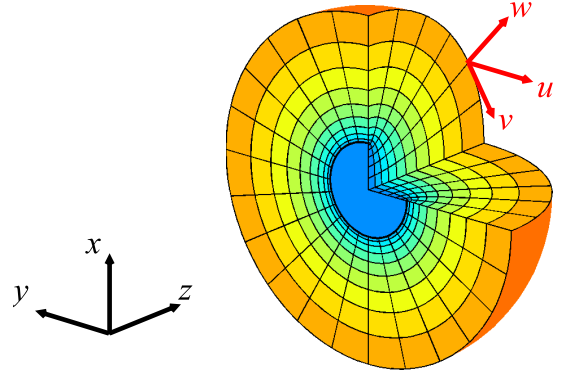}}
\caption{(Color online) The oblate spheroidal coordinate system for
  $0\le u\le 3\pi/2$ and $0\le v\le\pi/2$. Black lines are
  coordinate lines, and each color denotes a region between two
  constant-$w$ surfaces.}
\label{oblate_spheroids}
\end{figure}

The spherical lens is inconvenient for imaging and lithography, as the
object or the photoresist must be close to the inner surface of the
lens and must therefore also be spherical in shape. To make the object
surface flat, the oblate spheroidal coordinate system
\cite{arfken}, illustrated in
Fig.~\ref{oblate_spheroids}, is an ideal choice:
\begin{align}
x &= \alpha \cosh w \cos v \cos u, \nonumber\\
y &= \alpha \cosh w \cos v \sin u, \nonumber\\
z &= \alpha \sinh w\sin v ,\nonumber\\
h_u &= \alpha \cosh w \cos v , \nonumber\\
h_v &= h_w = \alpha\sqrt{ \sinh^2 w + \sin^2 v}.
\end{align}

We shall use the following transformation to map spheroidal
surfaces onto a single one:
\begin{align}
u &= u', &
v &= v', &
w &= \bigg\{\begin{array}{l}
w'+b,\\
b,\\
w',
\end{array}
&
\begin{array}{l}
0 \le w' < a,\\
a \le w' \le b,\\
w' > b,
\end{array}
\nonumber\\
\tilde{h}_{u'} &= 1, &
\tilde{h}_{v'} &= 1,&
\tilde{h}_{w'} &= \bigg\{\begin{array}{l}
1,\\
0,\\
1,
\end{array}
&
\begin{array}{l}
0 \le w' < a,\\
a \le w' \le b,\\
w' > b,
\end{array}
\end{align}
and let $a\to 0^+$ at the end of the calculation, so that the surface
$w' = a$ in the physical space becomes flat. Following the
coordinate transformation procedure, the desired physical
material constants are determined to be
\begin{align}
(\epsilon_{u'},\epsilon_{v'},\epsilon_{w'}) &=
\left(\epsilon_u\frac{\sinh^2b+\sin^2v'}{\cosh b\sin^2v'},
\epsilon_v\cosh b,\epsilon_w\cosh b\right),
\nonumber\\
(\mu_{u'},\mu_{v'},\mu_{w'}) &=
\left(\mu_u\frac{\sinh^2b+\sin^2v'}{\cosh b\sin^2v'},
\mu_v\cosh b,\mu_w\cosh b\right)
\end{align}
for $w' = 0$,
\begin{align}
(\epsilon_{u'},\epsilon_{v'},\epsilon_{w'}) &=
(\mu_{u'},\mu_{v'},\mu_{w'}) = (0,0,\infty)
\end{align}
for $0 < w' \le b$,
and 
\begin{align}
(\epsilon_{u'},\epsilon_{v'},\epsilon_{w'}) &=
(\mu_{u'},\mu_{v'},\mu_{w'}) = (1,1,1)
\end{align}
for $w' > b$.

\begin{figure}[htbp]
\centerline{\includegraphics[width=0.48\textwidth]{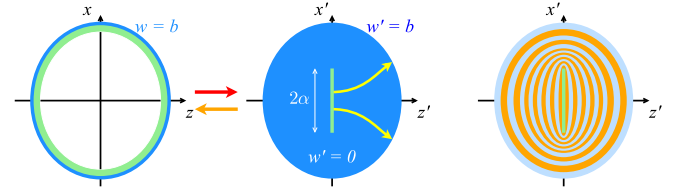}}
\caption{(Color online) Sketches of the oblate spheroidal perfect lens
  (center), the corresponding virtual space (left), and the
  metamaterial implementation of the lens (right). For clarity, only
  the $y' = 0$ cross sections are drawn. The structure is symmetric
  with respect to rotation about the $z'$ axis. Rays propagate along
  the $w'$ coordinate lines and follow hyperbolic trajectories.  In
  practice, one can use just half of the spheroidal lens ($z'>0$) and
  put the object against the $w'=0$ plane.}
\label{spheroidal}
\end{figure}

Figure \ref{spheroidal} sketches the geometry of the oblate spheroidal
lens.  Much like the previous examples, the lens consists of a highly
anisotropic material with zero transverse material constants and
infinite longitudinal constants, which can be implemented by thin
layers of oblate spheroidal films with alternate signs of permittivity
and permeability, as shown in Sec.~\ref{implement}. In this case, the
thicknesses of the films, $d_1$ and $d_2$, should be measured in terms
of the $w'$ coordinate.

Again, if the material at $w' = 0$ is made free space for practical
reasons, there will be impedance mismatch across the $w'=0$ interface
between the object plane and the spheroidal lens.  Once the fields
gets inside the metamaterial, however, the image at the plane $w'=0^+$
is perfectly magnified and transferred to free space for $w'>b$,
since the lens and the outer free space are perfectly matched layers.

\begin{figure}[htbp]
\centerline{\includegraphics[width=0.35\textwidth]{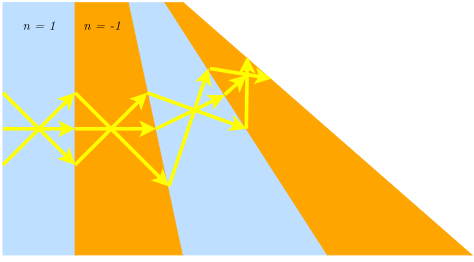}}
\caption{(Color online) A stack of slanted negative-index thin films
  can continuously redirect a ray with respect to the normal direction
  of each interface.}
\label{curved_stack}
\end{figure}

An interesting feature of the spheroidal lens is that the $w'$
coordinate lines are hyperbolic, so rays inside the transformation
medium are also hyperbolic and curved in general. Intuitively, the
curved rays can be understood in terms of negative refraction in the
ray optics picture, as shown in Fig.~\ref{curved_stack}, if the
negative-index thin film implementation is adopted. Negative
refraction can focus a point source in free space on the opposite side
of the interface \cite{pendry_prl,veselago}, so a stack of curved
negative-index thin films can continuously redirect a ray with respect
to the normal direction of each interface, causing the ray to be
curved.

\section{\label{superlens}Superlens Design}
The difficulty of controlling permeability without introducing
significant loss at optical frequencies has led researchers to the
concept of superlens, which has exotic permittivity values but unit
permeability and applies only to TM waves
\cite{pendry_prl,fang,melville,salandrino,jacob,liu,smolyaninov}.
Because the propagation of TM waves depends not only on the
permittivity tensor, but also the transverse permeability, one cannot
simply apply the perfect lens specifications on the permittivity only
and expect the metamaterial to behave like a perfect lens for TM
waves.  Instead, it is necessary to examine the TM wave propagation
behavior in such a material in order to determine the optimal
permittivity values, using the perfect lens design only as a
guideline.

To investigate superlensing in more general geometries, let us
consider the normalized Maxwell equations in arbitrary orthogonal
coordinates, given by Eqs.~(\ref{norm_maxwell}), inside a
superlens. Considering TM waves with nonzero $\tilde{H}_v$ only,
\begin{align}
\tilde{E}_{v} &= 0, 
\\
\tilde{H}_{u} = \tilde{H}_{w} &= 0, 
\\
\parti{\tilde{E}_u}{v} = \parti{\tilde{E}_{w}}{v} &= 0,
\\
\parti{}{v}\left(\tilde\mu_v\tilde{H}_v\right) &= 0,
\label{v_invariant}
\end{align}
Eqs.~(\ref{norm_maxwell}) become
\begin{align}
\parti{\tilde{E}_{u}}{w}-\parti{\tilde{E}_{w}}{u} &=
i\omega\mu_0 \tilde{\mu}_{v}\tilde{H}_{v},
\nonumber\\
\parti{\tilde{H}_{v}}{w} &=
i\omega\epsilon_0\tilde{\epsilon}_{u}\tilde{E}_{u},
\nonumber\\
\parti{\tilde{H}_{v}}{u} &= -i\omega\epsilon_0\tilde{\epsilon}_{w}\tilde{E}_{w}.
\label{tm_maxwell}
\end{align}
The analysis of TM waves with nonzero $\tilde{H}_u$ is similar.  The
wave equation in terms of $\tilde{H}_v$ is
\begin{align}
\left[\parti{}{w}\left(\frac{1}{\tilde{\epsilon}_{u}}\parti{}{w}
\right)+\parti{}{u}\left(\frac{1}{\tilde{\epsilon}_{w}}\parti{}{u}
\right)\right]\tilde{H}_{v} &=
-\frac{\omega^2}{c^2}\tilde{\mu}_{v}\tilde{H}_{v}.
\label{tm_wave}
\end{align}
If we make $\tilde\epsilon_{u} = 0$ as suggested by Salandrino and
Engheta \cite{salandrino}, the wave equation yields
$\partial\tilde{H}_v/\partial w = 0$, and the normalized magnetic
field is uniform with respect to $w$ inside the metamaterial. A point
source inside the metamaterial then produces a ray that
propagates in the $w$ direction. This phenomenon has been compared
\cite{salandrino} with resonance cones in plasma physics
\cite{bellan}. While such a propagation behavior resembles that in the
perfect lenses proposed in the previous sections, the zero transverse
permittivity causes significant impedance mismatch between the
metamaterial and free space, because resonance cones are well known to
be quasi-electrostatic waves and have an infinitesimal magnetic field
\cite{bellan}.

\begin{figure}[htbp]
\centerline{\includegraphics[width=0.48\textwidth]{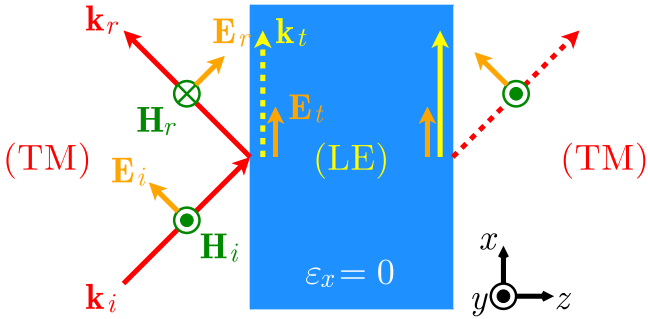}}
\caption{(Color online) TM waves cannot be coupled into a metamaterial
  slab with zero transverse permittivity, except when the plasmon
  resonance condition is met, because the TM waves inside the
  metamaterial are actually LE waves in most cases and have a zero
  magnetic field. By reciprocity, the LE waves also cannot be coupled
  into TM waves in free space.}
\label{TE_wave_mismatch}
\end{figure}

For example, consider the metamaterial slab with zero $\epsilon_x$
suggested in Refs.~\onlinecite{salandrino,ramakrishna}
(Fig.~\ref{TE_wave_mismatch}). In the Fourier domain, the wave
equation (\ref{tm_wave}) in Cartesian coordinates is
\begin{align}
\frac{k_z^2}{\epsilon_x} +\frac{k_x^2}{\epsilon_z} &= \frac{\omega^2}{c^2}.
\end{align} 
In the limit of $\epsilon_x\to 0$, $k_z$ is also zero.
Equations (\ref{tm_maxwell})  become
\begin{align}
- k_xE_z &= \omega\mu_0 H_y,
\nonumber\\
k_x H_y &= -\omega\epsilon_0\epsilon_z E_z,
\end{align}
If we requre $H_y$ to be nonzero,
\begin{align}
E_z &=-\frac{\omega\mu_0}{k_x} H_y =-\frac{k_x}{\omega\epsilon_0\epsilon_z} H_y,
\nonumber\\
k_x &= \sqrt{\epsilon_z}\frac{\omega}{c}.
\end{align}
Hence the TM wave inside the metamaterial can only have one specific
$k_x$.
%Using the thin film metamaterial implementation suggested in
%Refs.~\onlinecite{salandrino,ramakrishna} and outlined in
%Sec.~\ref{implement}, as well as the fact that
%\begin{align}
%\epsilon_{x} \approx
%\frac{\epsilon_1d_1+\epsilon_2 d_2}{d_1+d_2} = 0,
%\end{align}
%this specific $k_x$ is given by
%\begin{align}
%k_{x} &\approx
%\left(\frac{d_1+d_2}{d_1/\epsilon_1+d_2/\epsilon_2}\right)^{\frac{1}{2}}
%\frac{\omega}{c}
%\nonumber\\
%&= \left(\frac{1}{1/\epsilon_1+1/\epsilon_2}\right)^{\frac{1}{2}}
%\frac{\omega}{c},
%\end{align}
%which coincides with the surface plasmon polariton resonance
%condition for each interface between the thin films.

For any other $k_x$, $H_y$ and $E_z$ must vanish, and only $E_x$ can
be nonzero.  In other words, the waves become completely electric and
longitudinal, with the electric field parallel to the wave
vector. Since the magnetic boundary condition requires the magnetic
field to be continuous across the interface between the metamaterial
and free space, but the magnetic field of the longitudinal-electric
(LE) waves is zero, TM waves in free space cannot be coupled into the
LE waves inside the slab and must be completely reflected at the
boundary. By reciprocity, the LE waves, once excited inside the slab,
also cannot be coupled into free space at all. This means that the
planar superlens suggested in
Refs.~\onlinecite{ramakrishna,salandrino} is completely unable to
transmit an arbitrary TM image in and out of free space. This
impedance mismatch problem is especially severe for magnifying
superlenses, since one would be unable to observe the magnified image
in the far field, while any observed far-field radiation can only be
due to imperfections in the metamaterial implementation.

To partially overcome the impedance mismatch problem, it is more
desirable to make $\tilde\epsilon_{u}$ nonzero and
$\tilde\epsilon_{w} \to \infty$ instead. The wave equation in terms of
$\tilde{H}_v$ inside the metamaterial becomes
\begin{align}
\parti{}{w}\left(\frac{1}{\tilde\epsilon_u}\parti{\tilde{H}_v}{w}\right)
 &= -\frac{\omega^2}{c^2}\tilde{\mu}_{v}\tilde{H}_{v},
\label{tm_wave2}
\end{align}
and is independent of the transverse spatial profile of the fields.
The other conditions on the fields are
\begin{align}
\tilde{E}_{w} &= 0, \nonumber\\
\tilde{E}_u &=
\frac{1}{i\omega\epsilon_0\tilde\epsilon_u}\parti{\tilde{H}_v}{w},
\end{align}
resulting in transverse-electric-magnetic (TEM) waves, with a Poynting
vector in the $w$ direction regardless of the transverse spatial
profile.  Crucially, the transverse magnetic field is nonzero as long
as $\tilde\epsilon_u$ is also nonzero, allowing waves inside the
metamaterial to be partially coupled to TM waves in free space.  The
general solution of Eq.~(\ref{tm_wave2}) is
\begin{align}
\tilde{H}_v(u,v,w) = \tilde{H}_v(u,v,a)\tilde{W}_v(u,v,w),
\label{general_h}
\end{align}
where $\tilde{H}_v$ must satisfy Eq.~(\ref{v_invariant}) and
$\tilde{W}_v$ is the normalized magnetic field solution for a uniform
transverse spatial profile at $w = a$, that is, $\tilde{W}_v(u,v,a) =
1$ and $\tilde{W}_v$ satisfies
\begin{align}
\parti{}{w}\left(\frac{1}{\tilde\epsilon_u}\parti{\tilde{W}_v}{w}\right)
 &= -\frac{\omega^2}{c^2}\tilde\mu_v \tilde{W}_v.
\label{outgoing}
\end{align}
The boundary spatial profile $\tilde{H}_v(u,v,a)$ acts as a spatial
modulation of the field throughout propagation and does not diffract,
even though $\tilde{W}_v$ may change its shape along $w$. Thus, an
arbitrary TM image can be carried as a modulation of $\tilde{W}_v$
from one surface to another without loss of information.  For
lithography, the boundary spatial profile is applied at $w = b$, and
the converging wave solution of Eq.~(\ref{outgoing}) should be used
instead.

For instance, for a planar superlens with $\epsilon_z \to \infty$, we
obtain
\begin{align}
H_{y}(x,z) &= H_y(x,0)\exp
\left(i\sqrt{\epsilon_x} \frac{\omega}{c}z\right),
\nonumber\\
H_{x}(y,z) &= H_x(y,0)\exp
\left(i\sqrt{\epsilon_y} \frac{\omega}{c}z\right).
\end{align}
$k_z$ is constant, and a TM image can be perfectly transmitted inside
the lens, apart from an unimportant phase factor. Using the
approximate effective medium theory outlined in
Sec.~\ref{implement},
\begin{align}
\frac{d_1}{\epsilon_1}+\frac{d_2}{\epsilon_2} &= 0,
\nonumber\\
\epsilon_{z} &\approx \infty,
\nonumber\\
\epsilon_{x} = \epsilon_{y} &\approx \epsilon_1 + \epsilon_2.
\end{align}
To make the waves propagating, $\epsilon_{x}$ and $\epsilon_{y}$ must
be positive.

Depending on loss and other limitations in the metamaterial
implementation, such as finite thicknesses of the thin films,
$\tilde\epsilon_w$ obviously cannot be infinite in practice.  In the
planar geometry, assuming $k_y = 0$ for simplicity, $k_z$ becomes
\begin{align}
k_z &= \sqrt{\epsilon_x}\sqrt{\frac{\omega^2}{c^2}
-\frac{k_x^2}{\epsilon_z}}
\nonumber\\
&=\sqrt{\epsilon_x}\frac{\omega}{c}
\left[1-\frac{1}{\epsilon_z}
\left(\frac{k_x\lambda}{2\pi}\right)^2\right]^{1/2},
\end{align}
where $\lambda$ is the free-space wavelength.  The Abbe limit for the
superlens is therefore roughly given by
\begin{align}
\Delta_{\textrm{min}} \sim \frac{\lambda}{2\sqrt{|\epsilon_z|}}.
\end{align}
The resolution limit depends directly on the magnitude of the
longitudinal refractive index $\sqrt{|\epsilon_z|}$.

Let us estimate the resolution limit at $\lambda = 365$ nm due to loss
in a stack of infinitesimally thin silver ($\epsilon_1 \approx
-2.4+0.25i$) and aluminium oxide ($\epsilon_2 \approx 3.2$) layers.
Using the approximate effective medium theory, the maximum
longitudinal index $\sqrt{|\epsilon_z|}$ is about 7.4, at a $d_1/d_2$
ratio of $0.75$. This means that the free-space resolution limit can
be beaten by roughly a factor of 7. To obtain a more accurate
assessment of the resolution limit and that in other geometries, more
numerical and experimental studies are needed.

The $\epsilon_w \to\infty$ condition can naturally be applied to
magnifying configurations. For spherical coordinates, the physical
solution of Eq.~(\ref{tm_wave2}) is the spherical wave,
\begin{align}
H_{\phi}(\theta,r) &= H_{\phi}(\theta,a)\frac{a}{r}\exp
\left[i\sqrt{\epsilon_\theta} \frac{\omega}{c}(r-a)\right],
\nonumber\\
H_{\theta}(\phi,r) &= H_{\theta}(\phi,a)\frac{a}{r}\exp
\left[i\sqrt{\epsilon_\phi} \frac{\omega}{c}(r-a)\right].
\end{align}
For oblate spheroidal coordinates, the spheroidal wave
functions are much more complicated and given by
\begin{align}
&\quad\parti{}{w}\left\{\frac{\cosh w}{\sinh^2 w+\sin^2 v}
\parti{}{w}\left[\sqrt{\sinh^2w+\sin^2v}H_v(u,v,w)\right]\right\}
\nonumber\\
&=-\frac{\omega^2}{c^2}\alpha^2\epsilon_u \cosh w
\sqrt{\sinh^2w+\sin^2v}H_v(u,v,w),
\nonumber\\
&\quad \parti{}{w}\left\{\frac{1}{\cosh w}\parti{}{w}
\left[\cosh w H_u(v,w)\right]\right\}
\nonumber\\
&=-\frac{\omega^2}{c^2}\alpha^2\epsilon_v
\left(\sinh^2w+\sin^2v\right)H_u(v,w).
\label{spheroidal_wave}
\end{align}
but arbitrary TM images can still be transmitted as transverse
spatial modulations of the spheroidal wave functions.

In the limit of high magnification, TM waves in free space become
approximately TEM waves, so the TEM waves inside the magnifying
superlenses can be efficiently coupled to free space, if $\epsilon_u$
is close to $1$.

\section{Conclusion}
In conclusion, we have outlined the procedure of magnifying perfect
lens and superlens design by the coordinate transformation
technique. The use of oblate spheroidal coordinates is especially
promising for subwavelength microscopy and lithography, as they
provide a more convenient flat object or image plane and enable
two-dimensional magnification beyond the diffraction limit. For a
simpler experimental setup, the elliptic cylindrical coordinates
\cite{arfken} can also be used to provide a flat object plane and
one-dimensional magnification. Given the recent success in superlens
experiments, the oblate spheroidal or elliptic cylindrical superlens
should be relatively straightforward to demonstrate experimentally.
Loss is a major problem, and more theoretical, numerical, and
experimental analysis is needed to evaluate the impact of loss and
other deviations from the ideal design in practice. In applications
where a strong signal is preferred and loss in metamaterials is a
major detrimental factor, resonantly-enhanced near-field imaging by
low-loss dielectric structures may be a better option \cite{tsang}.

\section*{Acknowledgments}
This work is supported by the DARPA Center for Optofluidic Integration
and the National Science Foundation through the Center for the Science
and Engineering of Materials (DMR-0520565).

\end{document}